\documentclass[a4paper,twocolumn]{esapub2005}

\input{epsf}

\def\CQG{{\it Class. Quantum Gravity} }

\def\MPL{{\it Mod. Phys. Lett.} }
\def\MNRAS{{\it Mon. Not. R. Ast. Soc.} }

\def\PR{{\it Phys. Rev.} }
\def\PRL{{\it Phys. Rev. Lett.} }

\def\ep{\epsilon}   
\def\th{\theta}

\def\om{\omega}  \def\De{\Delta} 
   
 \def\Om{\Omega}

 \def\frac#1#2{{\textstyle{{#1}\over
{#2}}}} 
\def\lsim{\mathrel{\rlap{\lower4pt\hbox{\hskip1pt$\sim$}}
\raise1pt\hbox{$<$}}}
\def\gsim{\mathrel{\rlap{\lower4pt\hbox{\hskip1pt$\sim$}}
\raise1pt\hbox{$>$}}} \def\sqr#1#2{{\vcenter{\vbox{\hrule height.#2pt
\hbox{\vrule width.#2pt height#1pt \kern#1pt \vrule width.#2pt} \hrule
height.#2pt}}}}

\def\beq{\begin{equation}} \def\eeq{\end{equation}}
\def\beqa{\begin{eqnarray}} \def\eeqa{\end{eqnarray}}

\long\def\symbolfootnote[#1]#2{\begingroup%
\def\thefootnote{\fnsymbol{footnote}}\footnote[#1]{#2}\endgroup}

\begin{document}

\title{The Galileo satellite constellation and Modifications to the Inverse-Square Law for Gravitation}

\author{J. P\'aramos$^{1}$}
\author{O. Bertolami$^{2}$\footnote{Also at Departamento de F\'isica, Instituto Superior T\'ecnico}}

\affil{Instituto de Plasmas e Fus\~ao Nuclear, \\Av. Rovisco Pais 1, 1049-001 Lisboa, Portugal\\
$^{1}$ e-mail address: paramos@ist.edu\\
$^{2}$ e-mail address: orfeu@cosmos.ist.utl.pt}

\date{\today}

\maketitle


\begin{abstract}
We consider the impact of a power-law correction to the Newtonian potential, inspired by ungravity or $f(R)$ extensions of the Standard Model, and draw conclusions on the
possibility of measuring the relevant parameters through observables made available by the
Galileo satellite positioning system.
\end{abstract}


\section{Introduction}

The Galileo positioning system poses a great opportunity, not only for
the improvement and development of new applications in navigation monitoring and
related topics, but also possibly for fundamental research in physics. Indeed,
together with the already deployed Global Positioning System (GPS) and Glonass,
satellite navigation
may be considered the first practical application where relativistic
effects are taken into account, not from the usual experimental scientific point of
view, but as a regular engineering constraint on the overall design
requirements. Indeed, effects arising from special and general relativity (GR) -- gravitational blueshift, time dilation and Sagnac
effect -- may account to as much as $\sim 40~\mu s/day$, which is many
orders of magnitude above the accuracy of the onboard clock deployed
in these systems. Moreover, the gravitational Doppler effect, of the
order of $V_N /c^2 \simeq 10^{-10}$ (where $V_N = GM_E /R_E$ is the
Newtonian potential, $G$ is Newton's constant, $M_E \simeq 6.0 \times
10^{24}~kg$ is the Earth's mass, $R_E \simeq 6.4 \times 10^6~m$ is
its radius and $c$ is the speed of light) falls within the $10^{-12}$
frequency accuracy of current space-certified clocks, and must
therefore be taken into account. In the GPS, this is done by imposing an
offset in the onboard clock frequency, while in the Galileo Navigation Satellite System (GNSS)
this correction should be accounted by the receiver. For further details,
the reader is directed to Refs. \cite{Ashby,Pascual,Rovelli,Bahder} and references
therein.

This said, it is not clear as to what extent the accuracy
of the Galileo positioning system may be improved -- which is designed to offer pinpoint localization within
an error margin of $1~m$, against the $10~m$ margin of previous the
GPS -- so to provide clues to the nature of models beyond the current gravitational physics
knowledge (see Refs. \cite{review,Will} for updated surveys).
In this study, we aim to establish bounds on the
detectability of extensions to the recently discussed unparticle extension of the Standard Model of particle interactions 
\cite{georgi} that manifest themselves through power-law corrections to the Newtonian potential \cite{GN}. These corrections are 
associated to the exchange of modes that do not correspond to particle states, usually referred to as {\it unparticles}. 
The exchange of spin-2 states lead to a correction to Newton's potential usually dubbed as {\it ungravity}.

This paper is organized as follows: firstly, we assess the main relativistic effects that are present in the GNSS. 
We proceed and consider the possibility of measuring the discussed corrections using the GNSS.

\section{Main relativistic effects}

\subsection{Frame of reference}

Assuming that all time-dependent effects are of cosmological origin, and
hence of order $H_0^{-1}$, where $H_0$ is Hubble's constant, one may
discard these as too small within the timeframe of interest; hence, one
assumes a static, spherically symmetric scenario, posited by the
standard Schwarzschild metric. In isotropic form, this is given by the
line element

\beqa ds^2 &=& -\left(1 + {2V\over c^2} \right)(c~dt)^2 + {1 \over 1+ {2V
\over c^2} } dV  \\ \nonumber && \cong -\left(1 + {2V\over c^2} \right)(c~dt)^2 +
\left(1- {2V \over c^2} \right) dV ~~,\eeqa

\noindent where $dV = dr^2 + d\Om^2$ is the volume element, and $V$ is
the gravitational potential. In the standard GR scenario, the latter
coincides with the Newtonian potential $V = V_N = -(1 +
\Sigma^n_{i=1}J_n) GM_E /r $, where the $J_n$ multipoles account for the effect
of geodesic perturbations and density profiles.

To this picture, one must introduce the rotation of the Earth with respect to
this fixed-axis reference frame, with angular velocity $\om= 7.29
\times 10^{-5}~rad/s$; by doing a coordinate shift $t' =t$, $r' = r$,
$\th'=\th$ and $\phi' = \phi - \om t'$, one gets the Langevin metric,
given by the line element

\beqa ds^2 &=& -\left[1 + {2V\over c^2} - \left({\om r \sin \th \over
c}\right)^2 \right](c~dt)^2 \\ \nonumber && + 2 \om r^2 \sin^2\th d\phi dt+\left(1 +
{2V \over c^2}\right) dV~~,\eeqa

\noindent where, for simplicity, primes were dropped. Asides from a
non-diagonal element, one obtains an addition to the gravitational
potential, which could be viewed as a centrifugal contribution due to
the rotation of the reference frame. One can then define an effective
potential $\Phi = 2V - (\om r sin \th)^2$; the parameterization of the
Earth's geoid is obtained by taking the multipole expansion of $V$ up
to the desired order and finding the equipotential lines $\Phi
=\Phi_0$ (the latter being the value of $\Phi$ at the Equator), and
solving for $r(\th,\phi)$.

In the above line elements, the coordinate time coincides with the
proper time of an observer at infinity. However, since one wishes to
evaluate the ground to orbit clock synchronization, it is advantageous
to rewrite the metric in terms of a rescaled time coordinate, which
coincides with the proper time of clocks at rest on the surface of the
Earth; this is best implemented by resorting to the above-mentioned
geoid, since its definition as an equipotential surface $\Phi =
\Phi_0$ indicates that all clocks at rest with respect to it will beat at the same rate;
hence, rescaling the time coordinate according to $t \rightarrow (1
+\Phi_0 / c^2) t$, one gets the metric given by the line element

\beqa ds^2 &=& - \left[ 1 + {2 (\Phi-\Phi_0) \over c^2} \right] (c ~dt)^2
\\ \nonumber && + 2 \om r^2 sin^2\th d\phi dt + \left(1 - {2V\over c^2}
\right)d\Om~~. \eeqa

Finally, if one gets back to a non-rotating frame, the metric is given by
the line element

\beq ds^2 = - \left[ 1 + {2 (V-\Phi_0) \over c^2} \right] (c ~dt)^2 +
\left(1 - {2V\over c^2} \right)d\Om~~. \eeq

\subsection{Constant and periodic clock deviation}

One may now consider the difference between the time elapsed on the
ground and the satellite clock; keeping only terms of order $c^{-2}$,
one finds that the proper time increment on the moving clock is given
approximately by

\beq d\tau = ds / c = \left( 1 + {V-\Phi_0 \over c^2} -{v^2 \over 2
c^2} \right) dt~~.\eeq

\noindent
Considering an elliptic orbit with semi-major axis $a$, and
taking $V = V_N \simeq GM_E / r$, this may be recast into the form
\cite{Ashby}

\beqa && d\tau = ds / c = \\ \nonumber && \left[ 1 + {3GM_E \over 2 a c^2} + {\Phi_0 \over
c^2} - {2GM_E \over c^2} \left( {1 \over a } - {1 \over r}\right)
\right] dt~~.\eeqa

\noindent
The constant correction terms in the above amount
to \beq {3GM_E \over 2 a c^2} + {\Phi_0 \over c^2} = -4.7454 \times
10^{-10}~~,\eeq

\noindent
for the GNSS, and $-4.4647 \times 10^{-10}$, for
the GPS; this indicates that the orbiting clock is beating
faster, by about $41~\mu s /day$, for the GNSS, and
$39~\mu s /day$, for the GPS. For this reason, the GPS
has a built in frequency offset of this magnitude, while the increased
computational capabilities made available to current and future
receivers of the GNSS leave this correction to the user. The
residual periodic corrections, proportional to $1/r - 1/a$, have an
amplitude of order $49~ns/day$, for the GNSS, and
$46~ns/day$, for the GPS.

\subsection{Shapiro time delay and the Sagnac effect}

The so-called Shapiro time delay, a second order relativistic effect
due to the signal propagation is given by \cite{Ashby}

\beq \De t_{delay} = {\Phi_0 l \over c^3} + {2GM_E \over c^3}~ln
\left(1 + {l \over R_E} \right)~~,\eeq

\noindent where we have integrated over a straight line path of
(proper) length $l$. Evaluating this delay, one concludes that this
effect amounts to $6.67 \times 10^{-11}~s$.

Also, one must consider the so-called Sagnac effect, which arises from
the difference between the gravitational potential $V$ and the
effective potential $\Phi$, when proceeding from a non-rotational to a
rotational frame. Hence, one gets the additional time delay

\beq \De t_{Sagnac} = { \om \over c^2} \int_{path} r^2 ~d\phi = {2 \om
\over c^2} \int_{path} dA_z ~~, \eeq

\noindent
where $dA_z$ is the ortho-equatorial projection of the area
element swept by a vector from the rotation axis to the satellite. For
the GNSS, this yields a maximum value of $153~ns$, while for
the GPS, one gets $133~ns$.

One concludes this section by recalling the main effects affecting the
considered global positioning systems: a frequency shift of order
$ 10^{-10}$ and a propagation time delay (Shapiro plus Sagnac effect)
of the order $10^{-7}~s$. In what follows, one will compute the
additional frequency shift and propagation time delay induced by
ungravity inspired corrections to the Newtonian potential, and compare the results with
the above quantities, plus the frequency accuracy of $10^{-12}$ and
the time accuracy of Galileo, of order $10^{-9}~s$, which corresponds to a
optimistic spatial accuracy of $30~cm$.

\section{Detection of a power-law addition to the Newtonian potential}

Having studied the measurability of an array of hypothetical perturbations to the Newtonian potential in a 
previous work \cite{toulouse} (including the effect of the cosmological constant, of a Yukawa term \cite{Yukawa} 
and a constant acceleration such as the one reported in the so-called Pioneer anomaly --- see Ref. \cite{pioneer} 
and references therein for a review, and Ref. \cite{thermal} for a conventional explanation), one now considers a power-law perturbation of the form

\beq V_P = {G M_E \over r} \left( { R \over r} \right)^n~~, \label{power} \eeq

\noindent where $n$ is a (possibly non-integer) exponent and $R$ is a characteristic length scale arising from the underlying physical theory.

A power-law perturbation to the Newtonian poses an interesting scenario for several reasons: phenomenologically, it provides a convenient 
alternative to the more usual Yukawa parameterization of modifications to gravity, thus enabling one to probe potential candidates for 
extensions and modifications of GR. Furthermore, some current proposals give rise to the existence of induced power-law effects at 
astrophysical scales. As mentioned, corrections of this type arise from the exchange of spin-2 unparticles from which one can set bounds, for 
instance, from stellar stability considerations \cite{BPS09} and cosmological nucleosynthesis \cite{BS09}; in this context, the exponent is related 
to the scaling dimension of the unparticle operators $d_U$ through $n = 2d_U -2$, and the lengthscale $R$ reflects the energy scale of the unparticle 
interactions, the mass of exchange particles and the type of propagator involved.

Another relevant candidate for detection through a power-law parameterization is the so-called $f(R)$ class of theories \cite{fR}, 
which relies upon a generalization of the Einstein-Hilbert action (including a non-minimal coupling of geometry with matter \cite{f2}): 
in an astrophysical setting, these give rise to power-law additions to the Newtonian potential, which may be relevant in the context of the galaxy rotation puzzle \cite{capo,galaxy}.

In the above Eq. (\ref{power}), the Newtonian potential $\Phi_N$ is recovered by setting $R = 0$ (for positive $n$) or 
$R \rightarrow \infty$ (for negative $n$). The limit $n \rightarrow 0$ is ill-defined, since then the additional term $V_P$ does 
not vanish, but is identified with the Newtonian potential, $V_P = \Phi_N$: this indicates that one should rewrite the gravitational constant in terms of an effective coupling, leading to the full potential

\beq \Phi =  {G_P M_E \over r} \left[ 1 +  \left({ R \over r} \right)^n \right]~~,\eeq

\noindent with 

\beq G_P = {G \over 1 + \left({R \over R_0}\right)^n} ~~,\eeq

\noindent where $R_0$ signals the distance at which the full gravitational potential matches the Newtonian one, $\Phi(R_0) = \Phi_N (R_0)$.

This additional length scale $R_0$ should arise as an integration constant when solving the full field equations derived 
from the fundamental theory of gravity beyond the considered power-law correction. For simplicity, this study assumes that $(R /R_0)^n \ll 1$, so that this 
may be safely discarded --- at the cost of neglecting the regime $n \rightarrow 0$, for which this condition fails (notice that this is complementary to the 
approach followed in a previous study \cite{BPS09}). Hence, the subsequent section uses $G_P = G$ freely, albeit with the above caveat in mind.

\subsection{Relative frequency shift}

The relative frequency shift of a signal emitted at a distance from the origin $r=R_E+h$ (for the GNSS, $h = 23.222 \times 10^3~km$) and received at a distance $r=R_E$ is given by

\beqa \ep &=& 1 - {f_{Earth} \over f_{Sat} } = 1 - \sqrt{g_{00~Earth} \over g_{00~Sat}} = \\ \nonumber &&1 -  \sqrt{1 - 2V(R_E)/c^2 \over 1-2V(R_E+h)/c^2} \simeq \\ \nonumber &&  {V(R_E)-V(R_E+h) \over c^2}~~. \eeqa

\noindent Hence, one may compute the additional frequency shift induced by the extra contribution to the potential, through

\beqa \label{delta} \left({\De f \over f }\right)_P &=& {V_P(R_E) - V_P(R_E+h) \over c^2} \\ \nonumber & \simeq& {GM_E \over c^2 R_E} \left( {R\over R_E} \right)^n  \left[ 1 -  \left( {R_E\over R_E+h} \right)^{n+1} \right]  \\ \nonumber &=& 6.96 \times 10^{-7} \xi^n \left[ 1 - 0.22^{n+1} \right]~~, \eeqa

\noindent
where a dimensionless ratio $\xi \equiv R/R_E$ has been defined. One clearly has two asymptotic regimes: for $n \gg -1$, one may approximate the above by

\beq  \left({\De f \over f }\right)_P = 6.96 \times 10^{-7} \xi^n~~,  \eeq

\noindent so that, considering the accuracy $\ep_{f_r} = 10^{-12}$ of the Galileo constellation, one has $\xi^n \leq 1.44 \times 10^{-6}$. Conversely, if $n \ll -1$, one obtains

\beq  \left({\De f \over f }\right)_P = - 1.50 \times 10^{-7} (0.22 \xi)^n~~,  \eeq

\noindent which, upon comparison with the accuracy $\ep_{f_r}  = 10^{-12}$, yields  $ (0.22 \xi)^{-n} \geq 1.50 \times 10^5$. Since the {\it r.h.s.} is larger than unity, this imposes the rather strong bound $\xi > (0.22)^{-1} \simeq 4.64$ when $n\rightarrow -\infty$.

\begin{figure}[]
\centering
\epsfxsize= \columnwidth
\epsffile{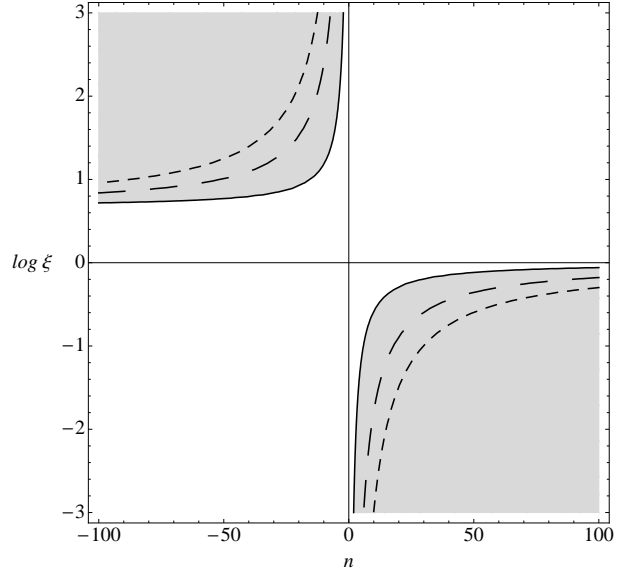}
\caption{Contour plot for the relative frequency deviation $\ep$ as a function of $\xi = R/R_E$ and $n$, with contours for $\ep = 10^{-12}$ (solid line), $ 10^{-24}$ (long dash) and $10^{-36}$ (short dash), and allowed region grayed out.}
\label{graphep}
\end{figure}

In the vicinity of $n =-1$, one expands Eq. (\ref{delta}) as

\beq  \left({\De f \over f }\right)_P = 1.07 \times 10^{-6} \left({n+1 \over  \xi}\right)~~,  \eeq

\noindent and the considered accuracy yields $\xi \geq 1.07 \times 10^6 |n+1|$. The different regimes are depicted in Fig. \ref{graphep}.

\subsection{Propagation time delay}

Besides the relative frequency shift discussed above, a power-law correction to the gravitational Newtonian potential also induces a modification to the propagational time delay, given by

\beqa \De t_P &=& {1 \over c} \int_{R_E}^{R_E+h} {GM_E \over r}\left({R \over r}\right)^n dr \\ \nonumber &=& {GM \over nc} \left( {R \over R_E}\right)^n \left[ 1 - \left({R_E \over R_E + h}\right)^n \right] \\ \nonumber &\simeq& 1.33 \times 10^6 {\xi^n \over n} \left( 1 - 0.22^n \right)  ~~. \eeqa

\noindent
Equating this to the $10^{-9}~s$ time resolution of the GNSS, yields the constraint 

\beq \xi^n \left| {1-0.22^n \over n} \right|  \leq 7.52 \times 10^{-16} ~~.\eeq 

\noindent 
Analogously to the previous discussion, for the extreme negative power-law regime $n \ll 0 $ this expression reads  $\left|0.22 \xi \right|^n  \leq 7.52 \times 10^{-16} |n| $. Since, for sufficiently large (negative) $n$, the {\it r.h.s.} is larger than unity, one recovers the stronger bound, $\xi > (0.22)^{-1} \simeq 4.64$, obtained above.

\begin{figure}[]
\centering
\epsfxsize= \columnwidth
\epsffile{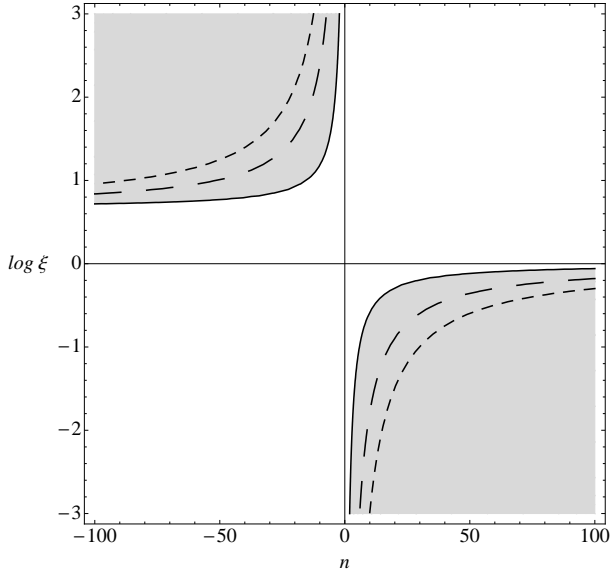}
\caption{Contour plot for the propagation time delay $\De t_P$ as a function of $\xi = R/R_E$ and $n$, with contours for $\De t_P = 10^{-9}~s $ (solid line), $ 10^{-12}~s$ (long dash) and $10^{-15}~s$ (short dash), and allowed region grayed out.}
\label{graphdt}
\end{figure}

 The allowed region for the $\xi$, $n$ parameters is depicted in Fig. \ref{graphdt}.

\section{Conclusions}

In this contribution, one has assessed the possibility of detecting signals of new physics through the use of the GNSS. This application could be valuable, as any unexpected new phenomenology could provide further insight into what lies beyond the Standard Model of particle interactions and GR. One has specifically looked at the propagation time delay and frequency shift induced by a putative correction to the Newtonian potential, namely one of the a power-law type.
This study complements the one where the GNSS was used to examine a constant contribution, cosmological constant induced and Yukawa-type corrections to the Newtonian potential \cite{toulouse}.

As can be seen from Figs. \ref{graphep} and \ref{graphdt}, one cannot lift the degeneracy in the $(R_E,n)$ parameter space through an intersection of the relative frequency shift and propagation time delay observables, since these produce rather similar allowed regions for the mentioned parameters.

Asides from the obtained range for these quantities, the main result of this study lies in the exclusion of a definite range of values for $R$, namely those lying between $0\leq R \leq 4.64 R_E$. Recall, however, that condition $(R /R_0)^n \ll 1$ leads to the limit $R_0 < R$ when $n \rightarrow -\infty$, yielding an additional requirement for the normalization constant $R_0$ in the case of a large negative exponent $n$. Also, the above results are not valid in the vicinity of $n=0$, due to the lack of knowledge concerning the effective gravitational coupling $G_P$.

\end{document}